\documentclass[prd,aps,twocolumn,amsmath,amssymb,nofootinbib,preprintnumbers]
{revtex4}

\usepackage{graphicx}
\usepackage{dcolumn}
\usepackage{bm}
\usepackage{amsmath}
\usepackage{amsthm}

\usepackage{amsfonts}
\usepackage{euscript,bbm}
\usepackage{ifthen}
\usepackage{psfrag}
\usepackage{slashed}

\def\ls{\mathrel{\lower4pt\vbox{\lineskip=0pt\baselineskip=0pt
           \hbox{$<$}\hbox{$\sim$}}}}
\def\gs{\mathrel{\lower4pt\vbox{\lineskip=0pt\baselineskip=0pt
           \hbox{$>$}\hbox{$\sim$}}}}
\def\drawbox#1#2{\hrule height#2pt

\hbox{\vrule width#2pt height#1pt \kern#1pt
              \vrule width#2pt}
              \hrule height#2pt}

\def\Asym#1#2{\vcenter{\vbox{\drawbox{#1}{#2}
              \kern-#2pt       
              \drawbox{#1}{#2}}}}

\newcommand{\be}{\begin{equation}}
\newcommand{\ee}{\end{equation}}
\newcommand{\bea}{\begin{eqnarray}}
\newcommand{\eea}{\end{eqnarray}}

\newcommand{\neu}[1]{\ensuremath{\tilde{\chi}_{#1}^0}}

\newcommand{\chpm}[1]{\ensuremath{\tilde{\chi}_{#1}^{\pm}}}
\newcommand{\chmp}[1]{\ensuremath{\tilde{\chi}_{#1}^{\mp}}}

\newcommand{\st}{\ensuremath{\tilde{t}}}

\newcommand{\gsim}{\lower.7ex\hbox{$\;\stackrel{\textstyle>}{\sim}\;$}}
\newcommand{\lsim}{\lower.7ex\hbox{$\;\stackrel{\textstyle<}{\sim}\;$}}

\newcommand{\tbar}{\overline{t}}

\newcommand{\met} {{E\!\!\!\!/_{\rm T}}}

\newcommand{\pT} {{p_{\rm T}}}

\newcommand{ \pgs }    {{\tt PGS4}}

\newcommand{ \madgraph } {{\tt MADGRAPH5}}

\begin{document}

%
\title{Probing Dark Matter at the LHC using Vector Boson Fusion Processes}


\author{Andres G. Delannoy$^{2}$}
\author{Bhaskar Dutta$^{1}$}
\author{Alfredo Gurrola$^{2}$}
\author{Will Johns$^{2}$}
\author{Teruki Kamon$^{1,3}$}
\author{Eduardo Luiggi$^{4}$}
\author{Andrew Melo$^{2}$}
\author{Paul Sheldon$^{2}$}
\author{Kuver Sinha$^{1}$}
\author{Kechen Wang$^{1}$}
\author{Sean Wu$^{1}$}

\affiliation{$^{1}$~Mitchell Institute for Fundamental Physics and Astronomy, \\
Department of Physics and Astronomy, Texas A\&M University, College Station, TX 77843-4242, USA \\
$^{2}$~Department of Physics and Astronomy, Vanderbilt University, Nashville, TN, 37235, USA \\
$^{3}$~Department of Physics, Kyungpook National University, Daegu 702-701, South Korea \\
$^{4}$~Department of Physics, University of Colorado, Boulder, CO 80309-0390, USA 
}

\begin{abstract}

Vector boson fusion (VBF) processes at the Large Hadron Collider (LHC) provide a unique opportunity to search for new physics with electroweak couplings. A feasibility study for the search of supersymmetric dark matter in the final state of two VBF jets and large missing transverse energy is presented at 14 TeV. Prospects for determining the dark matter relic density are studied for the cases of Wino and Bino-Higgsino dark matter. The LHC could probe Wino dark matter with mass up to approximately $600$ GeV with a luminosity of $1000$ fb$^{-1}$. 


\end{abstract}
MIFPA-13-15
\maketitle


Nearly $80\%$ of the matter of the Universe is dark matter (DM) \cite{WMAP}. The identity of DM is one of the most profound questions at the interface of particle physics and cosmology. Weakly interacting massive particles (WIMPs) are particularly promising DM candidates that can explain the observed relic density and are under investigation in a variety of direct and indirect searches. Within the context of $R$-parity conserving supersymmetric extensions of the standard model (SM), the WIMP DM candidate is the lightest supersymmetric particle (LSP), typically the lightest neutralino ($\neu{1}$), which is a mixture of Bino, Wino, and Higgsino states. 

The DM relic density is typically determined by its annihilation cross section at the time of thermal freeze-out. For supersymmetric WIMP DM, the annihilation cross section depends on the mass of $\neu{1}$ and its couplings to various SM final states, for which a detailed knowledge of the composition of $\neu{1}$ in gaugino/Higgsino states is required. Moreover, other states in the electroweak sector, such as sleptons, staus, or charginos can enter the relic density calculation. 

It is important to probe the electroweak sector of supersymmetric models directly in order to study their DM connection.  The main challenge to a direct probe of the electroweak sector at the Large Hadron Collider (LHC) is the small production cross section of neutralinos, charginos, and sleptons \cite{ATLASneutchargino}. 


In this Letter we explore supersymmetric DM produced directly at the LHC using vector boson fusion (VBF) processes \cite{Cahn:1983ip, Bjorken:1992er}. This appears particularly promising since some of the present authors recently showed that VBF production is quite effective in probing the chargino-neutralino system \cite{Dutta:2012xe}. It has also been suggested that VBF processes might be useful in both Higgs boson and supersymmetry studies \cite{Rainwater:1998kj, Choudhury:2003hq,cho,datta, Giudice:2010wb}. VBF production is characterized by the presence of two tagging jets with large dijet invariant mass in the forward region in opposite hemispheres. As shown in \cite{Dutta:2012xe}, the requirement of tagging jets along with missing transverse energy ($\met$) is very efficient in reducing SM background.

We also show in this Letter that information about production cross sections in VBF processes and the distribution of $\met$ in the final state can be used to solve for the mass and composition of $\neu{1}$, and hence the DM relic density. The cases of pure Wino or Higgsino $\neu{1}$, as well as the case of a mixed Bino-Higgsino $\neu{1}$ are studied. 

We note that the production of squarks ($\tilde{q}$) or gluinos ($\tilde{g}$) through gluon fusion, followed by cascade decay ending in the production of DM, is the classic setting for DM searches in final states with appreciable missing energy, multiple jets and leptons. However, determining the content of the neutralino and the masses of the superpartners without any color charges requires specific model dependent correlation between masses of colored and non-colored superpartners. In very specific settings, it is possible to determine the composition of $\neu{1}$ \cite{content}, as well as the mass of light staus or sleptons \cite{relic}. In general, the combinatoric background poses a major problem for such attempts. 

Recently, experiments at the 8-TeV LHC (LHC8) have put lower bounds on the masses of the $\tilde{g}$ and $\tilde{q}$. For comparable masses, the exclusion limits are approximately $1.5$ TeV at $95\%$ CL with $13$ fb$^{-1}$ of integrated luminosity \cite{:2012rz, Aad:2012hm, :2012mfa}. There are also active searches for the lightest top squark ($\tilde{t}$), and exclusion limits in the $m_{\st}$-$m_{\neu{1}}$ plane have been obtained in certain decay modes  \cite{ATLASStop1, ATLASStop2}. 

A direct probe of the electroweak sector using VBF processes is complementary to such searches. A variety of possibilities exist for the colored sector (compressed spectra, mildly fine-tuned split scenarios \cite{squarkheavy}, non-minimal supersymmetric extensions, etc.) with varying implications for existing and future searches. Experimental constraints (e.g. triggering) significantly affect the ability to probe supersymmetric DM in some of the above scenarios, for example those with compressed spectra. The important point to note is that \textit{a direct probe of the electroweak sector is largely agnostic about the fate of the colored sector} and provides a direct window to DM physics. 

The strategy pursued in this Letter will be to investigate direct DM production by VBF processes in events with $2j \, + \, \met$ in the final state. Such an approach has several advantages. The $2j \, + \, \met$ final state configuration provides a search strategy that is free from trigger bias. This is reinforced as the $p_T$ thresholds for triggering objects are raised by ATLAS and CMS experiments.
 




In order to probe DM directly, the following processes are investigated:
%
%
\be
pp \rightarrow \neu{1} \, \neu{1} \, jj, \,\,\, \chpm{1} \, \chmp{1} \, jj , \,\,\, \chpm{1} \, \neu{1} \, jj \,\,\,\,.
\ee
The main sources of SM background are: $(i)$ \,\, $pp \rightarrow Z jj \rightarrow \nu \nu j j $ and $(ii)$ \,\, $pp \rightarrow W jj \rightarrow l \nu j j $. The former is an irreducible background with the same topology as the signal. The $\met$ comes from the neutrinos. The latter arises from events which survive a lepton veto; $(iii)$ \,\, $pp \rightarrow t\tbar + $jets: This background may be reduced by vetoing $b$-jets, light leptons, $\tau$ leptons and light-quark/gluon jets.

The search strategy relies on requiring the tagged VBF jets, vetoes for $b$-jets, light leptons, $\tau$ leptons and light-quark/gluon jets, and requiring large $\met$ in the event. Signal and background events are generated with \madgraph \,\, \cite{Alwall:2011uj}. The detector simulation code used here is \pgs \, \cite{pgs}. 

Distributions of $\pT(j_1), \pT(j_2), M_{j_1j_2}$, and $\met$ for background as well as VBF pair production of DM are studied at $\sqrt{s}=8$ TeV and 14 TeV. In the case of pure Wino or Higgsino DM, $\chpm{1}$ is taken to be outside the exclusion limits for ATLAS' disappearing track analysis \cite{chargedtrack} and thus VBF production of $\chpm{1} \chpm{1}$, $\chpm{1} \chmp{1}$, and $\chpm{1} \neu{1}$ also contribute. The $\neu{1}$ masses chosen for this study are in the range 100 GeV to 1 TeV. The colored sector is assumed to be much heavier. There is no contribution to the neutralino production from cascade decays of colored particles. 

Events are preselected by requiring $\met > 50$ GeV and the two leading jets ($j_{1}$,$j_{2}$) each satisfying $p_{T} \geq 30$ GeV with $|\Delta \eta(j_{1},j_{2})| > 4.2$ and $\eta_{j_{1}}\eta_{j_{2}} < 0$. The preselected events are used to optimize the final selections to achieve maximal signal significance ($S / \sqrt{S + B}$). For the final selections, the following cuts are employed: $(i)$ The tagged jets are required to have $p_{T} > 50$ GeV and $M_{j_{1} j_{2}} > 1500$ GeV; $(ii)$ Events with loosely identified leptons ($l = e,\mu,\tau_{h}$) and $b$-quark jets are rejected, reducing the $t \tbar$ and $Wjj \rightarrow l\nu jj$ backgrounds by approximately $10^{-2}$ and $10^{-1}$, respectively, while achieving $99\%$ efficiency for signal events. The $b$-jet tagging efficiency used in this study is $70\%$ with a misidentification probability of $1.5\%$, following Ref. \cite{Chatrchyan:2012jua}. Events with a third jet (with $p_{T} > 50$ GeV) residing between $\eta_{j_{1}}$ and $\eta_{j_{2}}$ are also rejected; $(iii)$ The $\met$ cut is optimized for each different value of the DM mass. For $m_{\neu{1}} = 100$ GeV ($1$ TeV), $\met \geq 200$ GeV ($450$ GeV) is chosen, reducing the $Wjj\rightarrow l\nu jj$ background by approximately $10^{-3} \, (10^{-4})$. 

\begin{figure}[!htp]
\centering
\includegraphics[width=3.5in]{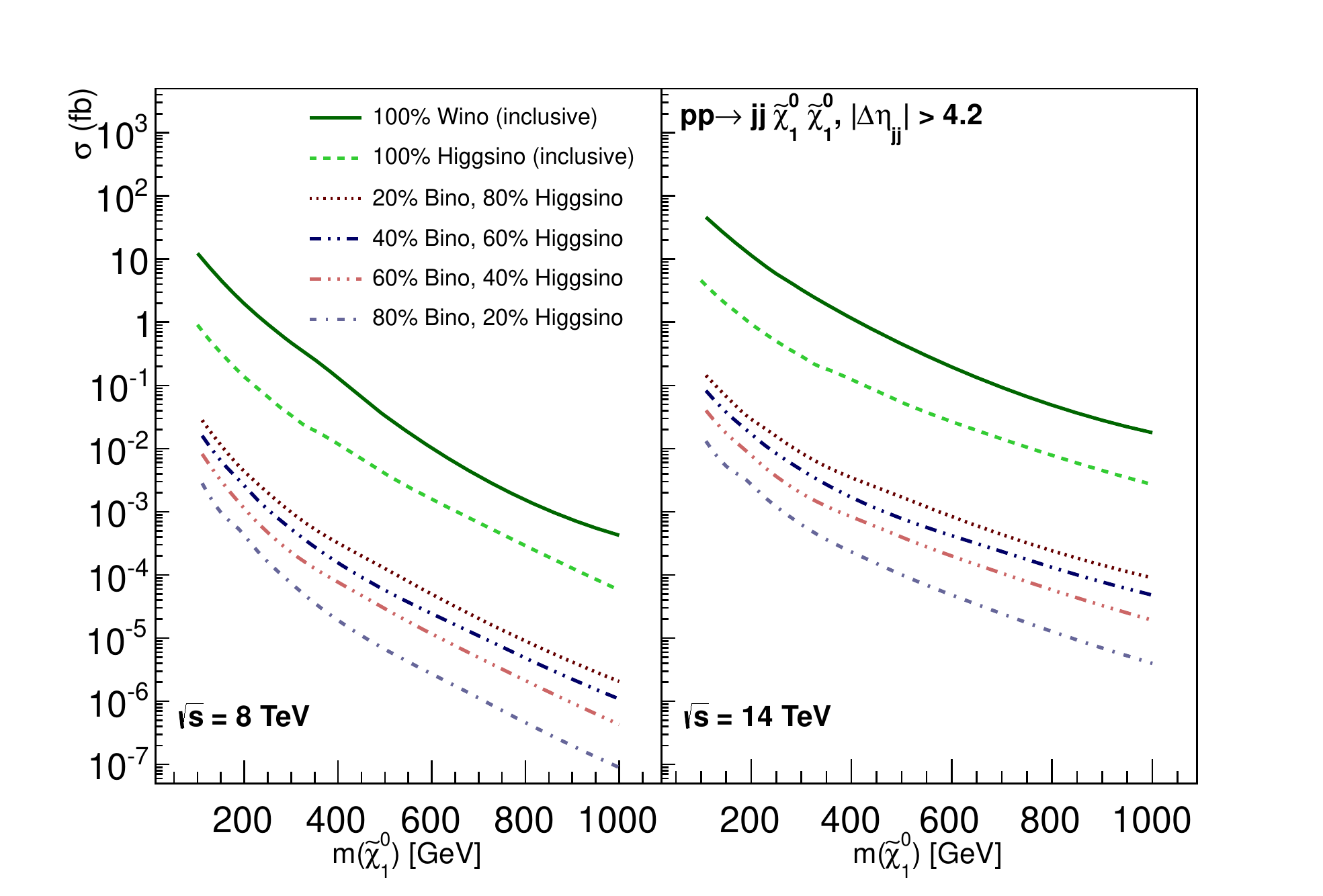}
\caption{Production cross section as a function of $m_{\neu{1}}$ after requiring $|\Delta \eta (j_1, j_2)| > 4.2$, at LHC8 and LHC14. For the pure Wino and Higgsino cases, inclusive $\neu{1} \neu{1}$,  $\chpm{1} \chpm{1}$, $\chpm{1} \chmp{1}$, and $\chpm{1} \neu{1}$ production cross sections are displayed.}
\label{xsectionColor}
\end{figure}

The production cross section as a function of $m_{\neu{1}}$ after requiring $|\Delta \eta (j_1, j_2)| > 4.2$ is displayed in Fig. \ref{xsectionColor}. The left and right panels show the cross sections for LHC8 and LHC14, respectively. For the pure Wino and Higgsino cases, inclusive $\neu{1} \neu{1}$,  $\chpm{1} \chpm{1}$, $\chpm{1} \chmp{1}$, and $\chpm{1} \neu{1}$ production cross sections are displayed. The green (solid) curve corresponds to the case where $\neu{1}$ is $99\%$ Wino. The inclusive production cross section is $\sim 40$ fb for a $100$ GeV Wino at LHC14, and falls steadily with increasing mass. The cross section is approximately $5-10$ times smaller for the pure Higgsino case, represented by the green (dashed) curve. As the Higgsino fraction in $\neu{1}$ decreases for a given mass, the cross section drops. For $20\%$ Higgsino fraction in $\neu{1}$, the cross section is $ \sim 10^{-2}$ fb for $m_{\neu{1}} = 100$ GeV at LHC14.

\begin{figure}[!htp]
\centering
\includegraphics[width=3.5in]{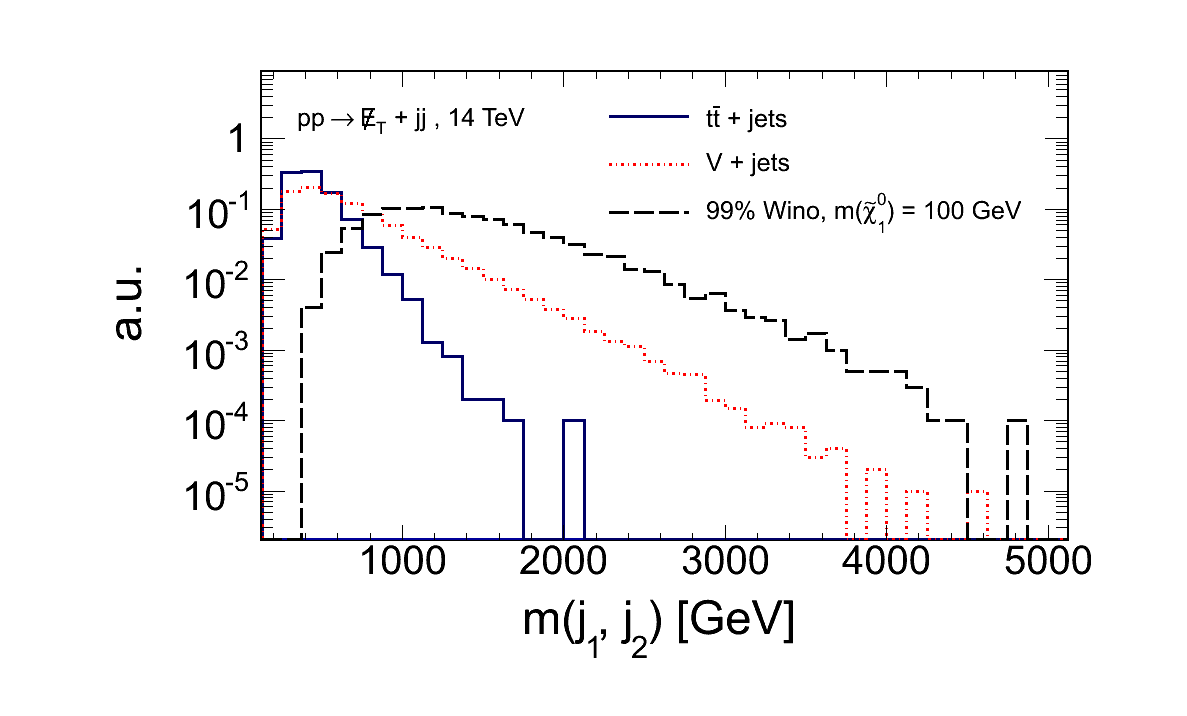}
\caption{Distribution of the dijet invariant mass $M_{j_1j_2}$ normalized to unity for the tagging jet pair $(j_1,j_2)$ and main sources of background after pre-selection cuts and requiring $p_T > 50$ GeV for the tagging jets at LHC14. The dashed black curves show the distribution for the case where $\neu{1}$ is a nearly pure Wino with $m_{\neu{1}} = 50$ and $100$ GeV. Inclusive $\neu{1} \neu{1}$,  $\chpm{1} \chpm{1}$, $\chpm{1} \chmp{1}$, and $\chpm{1} \neu{1}$ production is considered.}
\label{DiJetMass_VBFDM}
\end{figure}

Figure \ref{DiJetMass_VBFDM} shows the dijet invariant mass distribution $M_{j_1j_2}$ for the tagging jet pair $(j_1,j_2)$ and main sources of background, after the pre-selection cuts and requiring $p_T > 50$ GeV for the tagging jets at LHC14. The dashed black curves show the distribution for the case of a pure Wino DM, with $m_{\neu{1}} = 50$ and $100$ GeV.  The dijet invariant mass distribution for $W+$ jets, $Z+$ jets, and $t \bar{t} +$ jets background are also displayed. Clearly, requiring $M_{j_1j_2} > 1500$ GeV is effective in rejecting background events, resulting in a reduction rate between $10^{-4}$ and $10^{-2}$ for the backgrounds of interest.

\begin{figure}[!htp]
\centering
\includegraphics[width=3.5in]{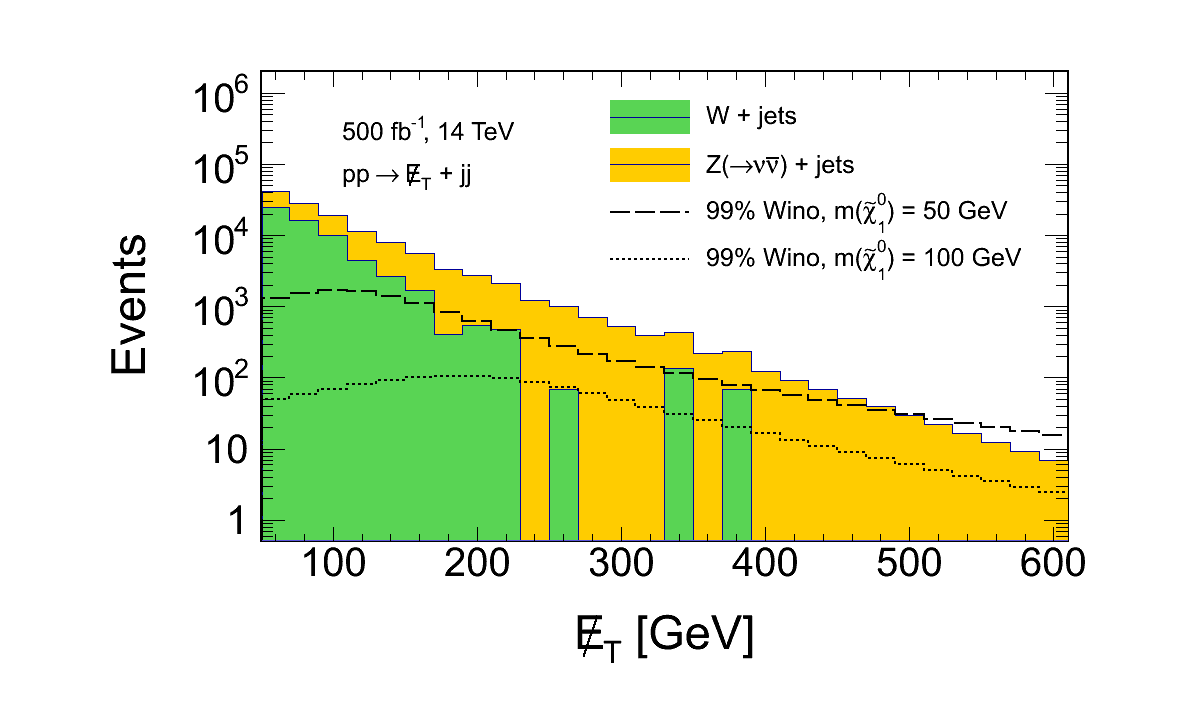}
\caption{The $\met$ distributions for Wino DM (50 GeV and 100 GeV) compared to $W+$ jets and $Z+$ jets events with $500$ fb$^{-1}$ integrated luminosity at LHC14. The distributions are after all selections except the $\met$ cut. Inclusive $\neu{1} \neu{1}$,  $\chpm{1} \chpm{1}$, $\chpm{1} \chmp{1}$, and $\chpm{1} \neu{1}$ production is considered.}
\label{Met_VBFDM_versionB}
\end{figure}

Figure \ref{Met_VBFDM_versionB} shows the $\met$ distribution for an integrated luminosity of 500 fb$^{-1}$ at LHC14  after all final selections except the $\met$ requirement. There is a significant enhancement of signal events in the high $\met$ region.


\begin{figure}[!htp]
\centering
\includegraphics[width=3.5in]{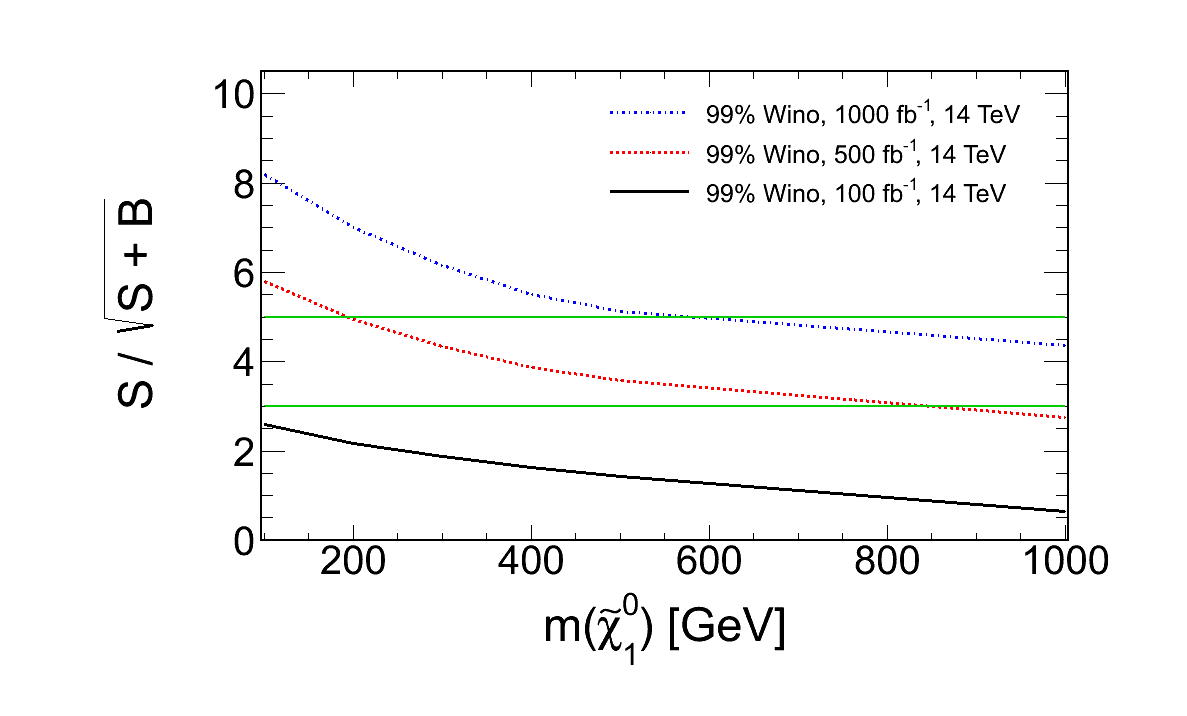}
\caption{Significance curves for the case where $\neu{1}$ is $99\%$ Wino as a function of $m_{\neu{1}}$ mass for different luminosities at LHC14. The green lines correspond to $3\sigma$ and $5\sigma$ significances.}
\label{SignificanceVsMassVsLumi_versionD}
\end{figure}

The significance as a function of $\neu{1}$ mass is plotted in Fig. \ref{SignificanceVsMassVsLumi_versionD} for different luminosities at LHC14. 
The blue, red, and black curves correspond to luminosities of $1000, 500,$ and $100$ fb$^{-1}$, respectively. At $1000$ fb$^{-1}$, a significance of $5\sigma$ can be obtained up to a Wino mass of approximately $600$ GeV.  

Determining the composition of $\neu{1}$ for a given mass is very important in order to understand early universe cosmology. For example, if $\neu{1}$ has a large Higgsino or Wino component, the annihilation cross section is too large to fit the observed relic density for $m_{\neu{1}}$ mass less than $\sim 1$ TeV for Higgsinos \cite{Allahverdi:2012wb} and $\sim 2.5$ TeV for Winos. On the other hand if $\neu{1}$ is mostly Bino, the annihilation cross section is too small. In the first case one has under-abundance whereas in the second case one has over-abundance of DM. Both problems can be solved if the DM is non-thermal \cite{Allahverdi:2012gk} (in the case of thermal DM, addressing the over abundance  problem requires addition effects like resonance, coannihilation etc. in the cross section, while the under-abundance problem can be addressed by having multi-component DM \cite{Baer:2012cf}). If  $\neu{1}$ is a suitable mixture of Bino and Higgsino, the observed DM relic density can be satisfied.   

From Figs. \ref{xsectionColor} and \ref{Met_VBFDM_versionB}, it is clear that varying of the rate and the shape of the $\met$ distribution can be used to solve for the mass of $\neu{1}$ as well as its composition in gaugino/Higgsino eigenstates. The VBF study described in this work was performed over a grid of input points on the $F - m_{\neu{1}}$ plane (where $F$ is the Wino or Higgsino percentage in $\neu{1}$). The $\met$ cut was optimized over the grid, and the $\met$ shape and observed rate of data were used to extract $F$ and $m_{\neu{1}}$ which was then used to determine the DM relic density. 



\begin{figure}[!htp]
\centering
\includegraphics[width=3.5in]{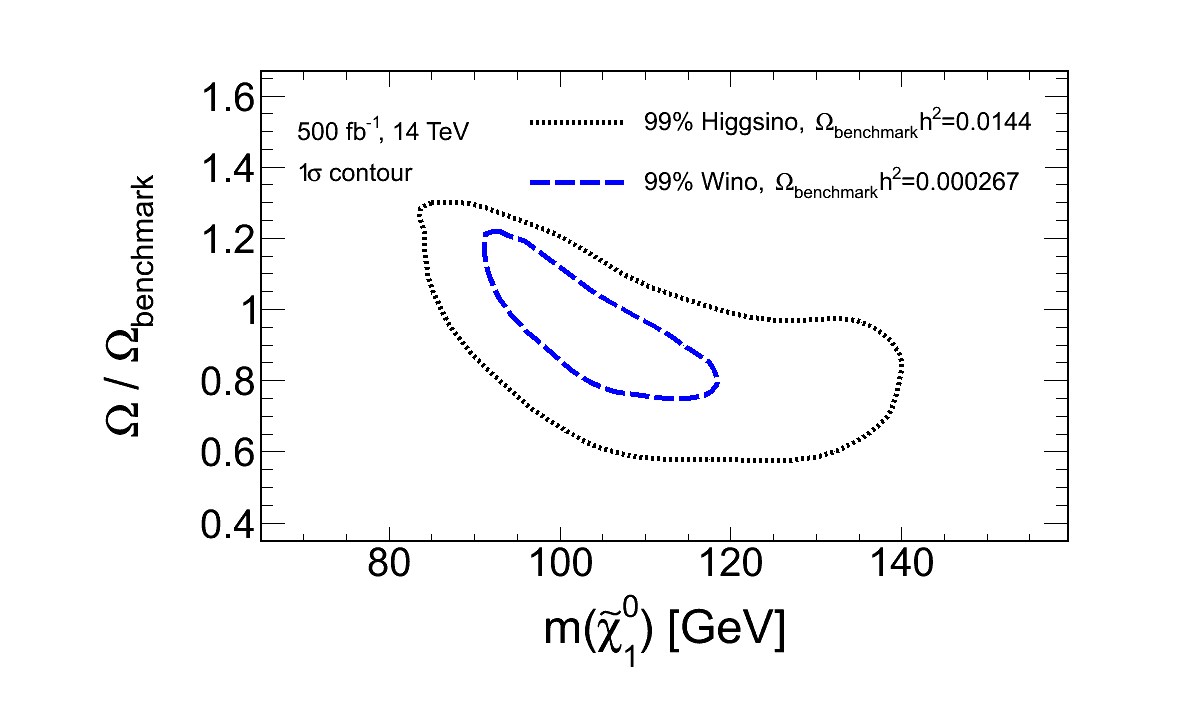}
\caption{Contour lines in the relic density-$m_{\neu{1}}$ plane for $99\%$ Wino (blue dashed) and $99\%$ Higgsino (grey dotted) DMs expected with $500$ fb$^{-1}$ of luminosity at LHC14. The relic density is normalized to its value at $m_{\neu{1}} = 100$ GeV.}
\label{OmegaVsMLSP_WinoAndHiggsino_versionD}
\end{figure}

In Fig. \ref{OmegaVsMLSP_WinoAndHiggsino_versionD}, the case of $99\%$ Higgsino and $99\%$ Wino were chosen, and $1\sigma$ contour plots drawn on the relic density-$m_{\neu{1}}$ plane for $500$ fb$^{-1}$ luminosity at LHC14. The relic density was normalized to a benchmark value $\Omega_{\rm benchmark}$, which is the relic density for $m_{\neu{1}} = 100$ GeV.  For the Wino case, the relic density can be determined within $\sim 20\%$, while for the Higgsino case it can be determined within $\sim 40\%$. For higher values of $m_{\neu{1}}$, higher luminosities would be required to achieve these results. We note we have not evaluated the impact of any degradation in $\met$ scale, linearity and resolution due to large pile-up events. Our results represent the best case scenario and it will be crucial to revisit with the expected performance of upgraded ATLAS and CMS detectors. 

In conclusion, this Letter has investigated the direct production of supersymmetric DM by VBF processes at the LHC. The cases of pure Wino, pure Higgsino, and mixed Bino-Higgsino DM have been studied in the $2j + \met$ final state at 14 TeV. The presence of the energetic VBF jets with large dijet invariant mass as well as the large $\met$ due to DM production have been used to reduce SM background. It has been shown that broad enhancements in the $\met$ and VBF dijet mass distributions provide a smoking gun signature for VBF production of supersymmetric DM. By optimizing the $\met$ cut for a given $m_{\neu{1}}$, one can simultaneously fit the $\met$ shape and observed rate in data to extract the mass and composition of $\neu{1}$, and hence solve for the DM relic density. At an integrated luminosity of $1000$ fb$^{-1}$, a significance of $5\sigma$ can be obtained up to a Wino mass of approximately $600$ GeV. The relic density can be determined to within $20\% \, (40\%)$ for the case of a pure Wino (Higgsino) for $500$ fb$^{-1}$ at LHC14, for $m_{\neu{1}} = 100$ GeV. We note that our study does not include the effect of large multiple interactions at high luminosity operations at the LHC. This is a very important subject, but outside the scope of the present work, because the final performance will depend on the planned upgrade of ATLAS and CMS detectors.



This work is supported in part by DOE Grant No. DE-FG02-95ER40917 and DE-FG02-04ER41290, NSF Award PHY-1206044, and
by the World Class University (WCU) project through the National Research Foundation (NRF) of Korea funded by the Ministry of Education, Science, and Technology (Grant No. R32-2008-000-20001-0). K.S. would like to thank Nathaniel Craig for helpful discussions.

\end{document}